\documentclass{article}

\usepackage{PRIMEarxiv}

\usepackage[utf8]{inputenc} 
\usepackage[T1]{fontenc}    
\usepackage{hyperref}       
\usepackage{url}            
\usepackage{booktabs}       
\usepackage{amsfonts}       
\usepackage{amsmath}
\usepackage{nicefrac}       
\usepackage{microtype}      
\usepackage{lipsum}
\usepackage{fancyhdr}       
\usepackage{graphicx}       
\usepackage{natbib}
\usepackage{eurosym}
\usepackage{booktabs,tabularx}

\DeclareMathOperator*{\argmax}{arg\,max}
\DeclareMathOperator*{\argmin}{arg\,min}

\title{Competition and Cooperation of Autonomous Ridepooling Services: Game-Based Simulation of a Broker Concept
\thanks{\textit{\underline{Submitted to}}: 
\textbf{Frontiers in Future Transportation}} 
}

\author{
  Roman Engelhardt, Patrick Malcolm, Florian Dandl, and Klaus Bogenberger \\
  Chair of Traffic Engineering and Control \\
  Technical University of Munich \\
  Arcisstraße 21 \\
  80333 Munich, Germany \\
  \texttt{Corresponding Author: roman.engelhardt@tum.de} \\
}

\begin{document}
\maketitle

\begin{abstract}
  With advances in digitization and automation, autonomous mobility on demand services have the potential to disrupt the future mobility system landscape. Ridepooling services in particular can decrease land consumption and increase transportation efficiency by increasing the average vehicle occupancy. Nevertheless, because ridepooling services require a sufficient user base for pooling to take effect, their performance can suffer if multiple operators offer such a service and must split the demand. This study presents a simulation framework for evaluating the impact of competition and cooperation among multiple ridepooling providers. Two different kinds of interaction via a broker platform are compared with the base cases of a single monopolistic operator and two independent operators with divided demand. In the first, the broker presents trip offers from all operators to customers (similar to a mobility-as-a-service platform), who can then freely choose an operator. In the second, a regulated broker platform can manipulate operator offers with the goal of shifting the customer-operator assignment from a user equilibrium towards a system optimum.
  To model adoptions of the service design depending on the different interaction scenario, a game setting is introduced. Within alternating turns between operators, operators can adapt parameters of their service (fleet size and objective function) to maximize profit. Results for a case study based on Manhattan taxi data, show that operators generate the highest profit in the broker setting while operating the largest fleet. Additionally, pooling efficiency can nearly be maintained compared to a single operator. With the resulting increased service rate, the regulated competition benefits not only operators (profit) and cities (increased pooling efficiency), but also customers. Contrarily, when users can decide freely, the lowest pooling efficiency and operator profit is observed.  
\end{abstract}


\section{Introduction}
With the increased availability of mobile internet, mobility-on-demand (MOD) services have become increasingly popular over the last decade. In times of urbanization, they can represent an alternative to private vehicles that offers a similar convenience. MOD services result in a higher temporal utilization of vehicles which, if replacing private vehicle trips, can potentially free up urban space that would otherwise be used for parking. Furthermore, ridepooling services have the potential to increase the average vehicle occupancy during trips, thereby resulting in more spatially efficient utilization of the road. The probability of finding and pooling similar trips increases with demand density. Therefore, the pooling potential increases with the scale of supply and demand. As a consequence, fragmentation of the ridepooling market into multiple independent competitors can be expected to decrease the efficiency of each competitor. Mobility-as-a-service (MaaS) platforms represent a possibility to break the stark independence of competitors, as offers from multiple mobility service providers are collected in one place for travelers.

Compared to current MOD services, automation can change the cost structure significantly~\cite{Bosch.2018}. When these cost reductions from the fleet operation with autonomous vehicles are translated into cheaper fares for users, disruptions of the transportation systems as we know them are possible. Hence, city authorities are confronted with the questions of whether and how autonomous mobility-on-demand (AMOD) systems and competition between multiple providers should be regulated.

\begin{figure}[htbp]
\begin{center}
\includegraphics[width=\textwidth]{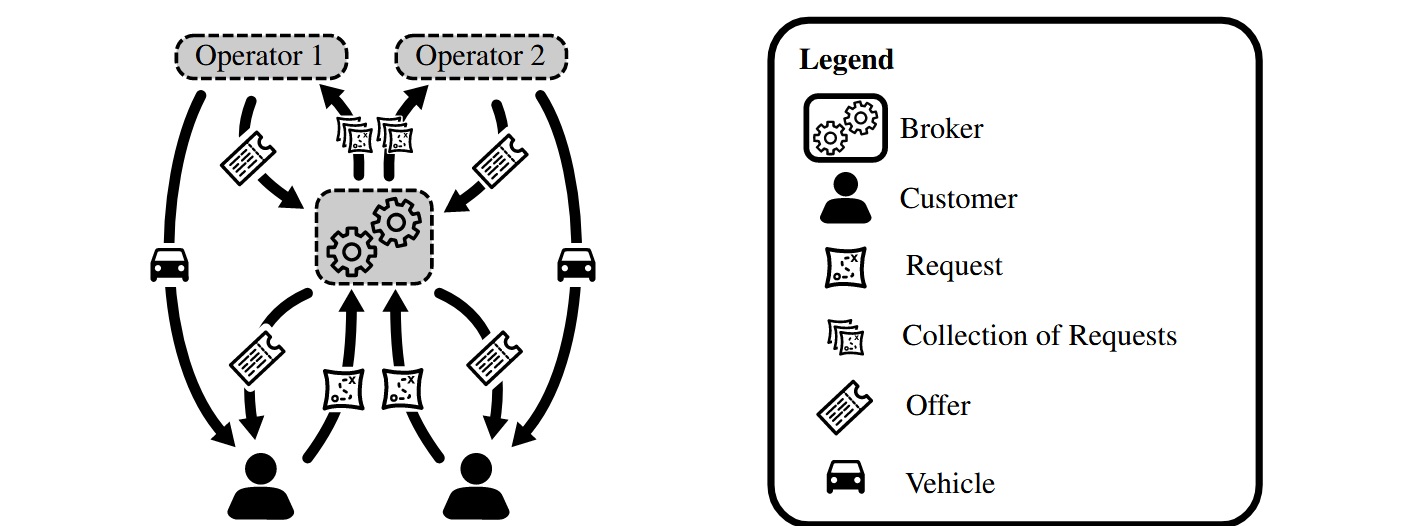}
\end{center}
\caption{Illustration of an AMOD Broker System}\label{fig:1}
\end{figure}

This paper studies how interaction between AMOD operators can counteract the effects of competition and fragmentation. To this end, the concept of an AMOD \textit{broker} is introduced which is a (possibly regulated) platform --- similar to a MaaS platform --- for multiple AMOD operators. As illustrated in Figure~\ref{fig:1}), the broker collects trip offers from multiple AMOD providers and forwards them to the customers. In addition, the broker can be regulated to adapt the offers to align the platform with city goals. The adaptation can range from sorting the offers in a certain order or manipulating prices, to the suppression of certain offers which are in conflict with city goals. The effect of the regulating measures can be compared with moving the dynamic traffic assignment from a user equilibrium towards the system optimum.

The goal of this study is to compare different types of AMOD provider interaction and their impact on the providers. More specifically, we investigate the following scenarios with the help of simulations: a monopolistic AMOD service, independent AMOD providers, and the two forms of broker systems (unregulated and fully regulated).

\subsection{Literature Review} 
Several studies deal with the operation of AMOD fleets and its impacts. Even without pooling, a single AV can replace a significant number of private vehicles~\cite{Fagnant.2015b} or carsharing vehicles~\cite{Dandl.2019b}. Optimization of request and repositioning assignments based on demand estimations can further improve fleet performance~\cite{Hyland.2018, Dandl.2019, Horl.2019}. However, without pooling, these vehicle reductions only affect stationary traffic, i.e. parking space. To observe improvements to traffic flow, ridepooling is required~\cite{Engelhardt.2019, Ruch.2020}. The optimization of ridepooling assignments is a challenging problem, which can be addressed with graph-based approaches~\cite{Santi.2014,AlonsoMora.2017} and heuristics based on them~\cite{Simonetto.2019,Hyland.2020}. For ridepooling services, positive scaling properties, i.e. a higher efficiency for higher levels of demand, are observed in both analytical and simulation models~\cite{Tachet.2017,Bilali.2020}. Two effects play into this scaling behavior for ridepooling systems: (i) a higher density of vehicles means that the approach becomes shorter (similarly to in the ridehailing case), and (ii) the probability of finding trips that can be matched with only minor detours increases with demand.

These operational studies assume a fixed exogenous demand, and the problem is to serve this demand as efficiently as possible. To study the impact of AMOD on transportation systems, the integration of AMOD into demand models is necessary. Open-source software packages like SimMobility~\cite{NahmiasBiran.2020,Oke.2020}, MATSim~\cite{Horl.2019b,Kaddoura.2020}, Polaris~\cite{Gurumurthy.2020}, and mobiTopp~\cite{Wilkes.2021b}, as well as commercial software solutions already have capabilities to model AMOD supply and demand interactions. Most of these demand models utilize a pre-day assignment of AMOD demand, be it by iterative learning or a mode choice model. \citet{Wilkes.2021b} developed a within-day mode choice model, which is based on real-time information of the fleet and thereby relevant for modeling MaaS platforms.

Most of the previously mentioned references study a single AMOD operator; a generalization to multiple independent operators has been implemented in \citet{Dandl.2019c}, and an operator with multiple service offers (hailing and pooling with different vehicle sizes) was investigated in~\cite{Atasoy.2015,Liu.2019a}. For ridehailing, the impact of multiple AMOD operators in the same market is analyzed with a theoretical model by \citet{Sejourne.2018} and data-driven models with simulation by \citet{Kondor.2022}. \citet{Sejourne.2018} show that demand patterns are crucial when it comes to the division of the market and find two phases. The first is denoted ``fragmentation resilient'' and describes a system where the price of sharing the market decreases with the size of the market; in the other phase, denoted ``fragmentation affected'', a division of the market generates much higher costs regardless of the size of the market because empty vehicle repositioning is required to balance supply and demand. \citet{Kondor.2022} derived a mathematical equation for the cost of non-coordinated market fragmentation and run simulations to find the coefficients for various cities. For ridepooling, \citet{Pandey.2019} analyzed three models of multi-company market models --- competitive, cooperative, and centralized --- and derived approaches to address the resulting problems with linear assignment problems.

There are several studies examining the even more complex market dynamics for multiple ridehailing services with drivers. For instance, \citet{Jiang.2018} study the effect of ``double-apping'' in a (human driven) ridehailing market with two competitors, where both drivers and customers have the possibility to use both ridehailing apps. They observed that users and drivers can benefit, but without any contract or guarantee that the other ridehailing provider will do the same, a ridehailing operator does not benefit from drivers serving customers of both providers. \citet{Qian.2017} study the competition between a traditional taxi and a ridehailing provider in a game, where passengers are the leaders and the two mobility providers are the followers. They find that fleet size and pricing policy significantly impact the outcome. In another study, \citet{Xi.2021} propose a ``name-your-own-price auction'' in a MaaS platform, where travelers and a wide range of mobility service providers (with ridehailing being one of them) can submit a bid. They also use a leader-follower formulation, with the MaaS platform being the leader and the mobility service providers and the travelers being the followers.

AMOD providers do not just compete against each other. They can compete with, but also complement public transport, depending on their service designs. In most demand models, AMOD systems are treated as a separate competing mode. Positive and negative effects mainly depend on the number of users that are attracted from private vehicle or public transport modes. Additionally, AMOD systems can be utilized as feeder systems to increase intermodality and improve public transport~\cite{Liang.2016,Wen.2018}. To avoid competition, AMOD routes can also be restricted/designed to complement the existing public transport system~\cite{Dandl.2019d}, or AMOD and public transport can be designed jointly~\cite{Pinto.2019b}. With a growing market share of today's MOD services, negative externalities of user-centric ridehailing can be observed~\cite{Henao.2019,Schaller.2021}. Therefore, the regulation of MOD~\cite{Li.2019b,Zhang.2019c} -- with part of the focus on the regulatory protection of drivers -- and of AMOD services~\cite{Simoni.2019,Dandl.2021b,Mo.2021} is becoming increasingly relevant. \citet{Simoni.2019} study various congestion pricing models in the presence of an AMOD system. \citet{Dandl.2021b} consider an AMOD ridepooling service which is regulated such that it cannot offer guaranteed single-passenger rides. Moreover, they introduce a regulatory tri-level framework optimizing a congestion-based road toll, parking fees, public transport frequency, and an AMOD fleet limit, where the reaction of an AMOD provider to changed regulatory settings is taken into account. \citet{Mo.2021} investigate how regulatory measures like fleet size limitations and public transport subsidies can steer the competition between AMOD and line-based public transport. The equilibrium state is found with an iterative approach, in which the AMOD operator is updated every iteration -- representing a day -- and the public transport service every month. These time scales should reflect the frequencies with which AMOD and public transport operators are likely to modify their service.

A collaboration of mobility services can help to create a better combined service offer, which could reduce private vehicle ownership and be beneficial to the service providers. MaaS platforms are one form of such collaboration. Typically, they at least collect information of multiple providers, offer the possibility to book mobility services, and provide a common method for payment~\cite{Smith.2020}. The design and possible regulation of a MaaS platform, e.g. by pricing and bundling of services, can affect user decisions~\cite{Feneri.2020} and ultimately help in reaching sustainability objectives~\cite{Muller.2021}.

\subsection{Contribution}
This paper contributes several new aspects to the literature. While most previous studies focused on the ridehailing market, this paper evaluates the losses resulting from fragmented ridepooling demand. Moreover, the effects of different interactions between multiple operators and a central platform are compared. The potential benefits of a broker which selects between the offers of different providers, thereby representing the most extreme form of regulation on this platform, is examined and compared to a platform where customers select the offers by themselves. To the authors knowledge, this is the first study that additionally evaluates the adoption of the service design to optimize the operators profit for a given interaction scenario within a game setting. The case study shows the significant impact that fleet size and the operator objectives have on the level of service and overall transportation system.

\section{Methodology}

This section describes the agent-based simulation environment, which is used to study different operator interactions. First, the simulation's agents and process flow are introduced, and the representation of different AMOD operators is explained. Then, the operator module with the task to assign customers to vehicles is described in detail. Lastly, an iterative simulation to model possible service adaptations to the studied operator interactions (independent, unregulated and regulated broker) is presented.

\subsection{Agent-Based Simulation Flow}
The simulation environment consists of three or four main agents: (1) customers, who request trips from AMOD operator(s) and choose their travel mode; (2) operators, who provide the mobility service by operating a fleet of vehicles with the tasks to create mobility offers for customer requests and fulfill these offers in case customers accept them; (3) vehicles controlled and dispatched by an operator which specifies which, where, and in which sequence customers have to be picked up and dropped off; and (4) a broker, which makes decisions to regulate the platform in the broker scenarios.

Customer and vehicle agents move on a network $G=(N,E)$ with nodes $N$ and edges $E$ connecting these nodes. A customer request is defined by the tuple $(i, t_i, x_i^s, x_i^d)$ with a request id $i$, a time of request $t_i$, the travel origin node of the request $x_i^s \in N$ and the travel destination node of the request $x_i^d \in N$. Operators receive these travel requests and, based on their current fleet state, try to find best possible solutions to serve them and formulate offers for the service as a reply. Offers from operators are defined as tuples of parameters defining the course of the customer's trip in case the offer is booked. In this study, parameters defining the offers can be categorized into user parameters $u_{i,o}$ and system parameters $s_{i,o}$, which influence the decision process of users and the broker, respectively. We define user parameters as parameters that users of the service are sensitive to when they have to decide for or against the service. These parameters can include fare, expected waiting time, and expected travel time for example. The broker on the other hand is sensitive to the system parameters. These parameters describe measures for the possible impact on the traffic system. In this study, the additional distance, which the AMOD fleet has to drive in order to serve a customer, is used. In case the operator is not able to serve a customer (i.e. no feasible solution is found to serve a customer within time constraints for pick-up and drop-off), no offer is made.

In so-called interaction scenarios, this study distinguishes several decision processes defining which specific operator is booked by a customer. The four different interaction scenarios implemented in this study are as follows:

\paragraph*{1) \textbf{Single Operator:}}
In this scenario, only a single monopolistic AMOD operator is offering a ridepooling service, and therefore no interaction between operators is implemented. Customers requesting a trip from this operator always book a trip if they receive an offer, and if not, they leave the system unserved.

\paragraph*{2) \textbf{Independent Operators:}}
In this scenario, multiple AMOD operators are offering a ridepooling service, but no direct interaction between them is assumed. Customers only request a trip from one of these operators, and they always book a trip if they receive an offer from this operator. If they don't receive an offer, they leave the system unserved. From the simulation point of view, this scenario is equivalent to the Single Operator scenario, but with the demand for AMOD being split between the operators.

\begin{figure}[htbp]
\begin{center}
\includegraphics[width=\textwidth]{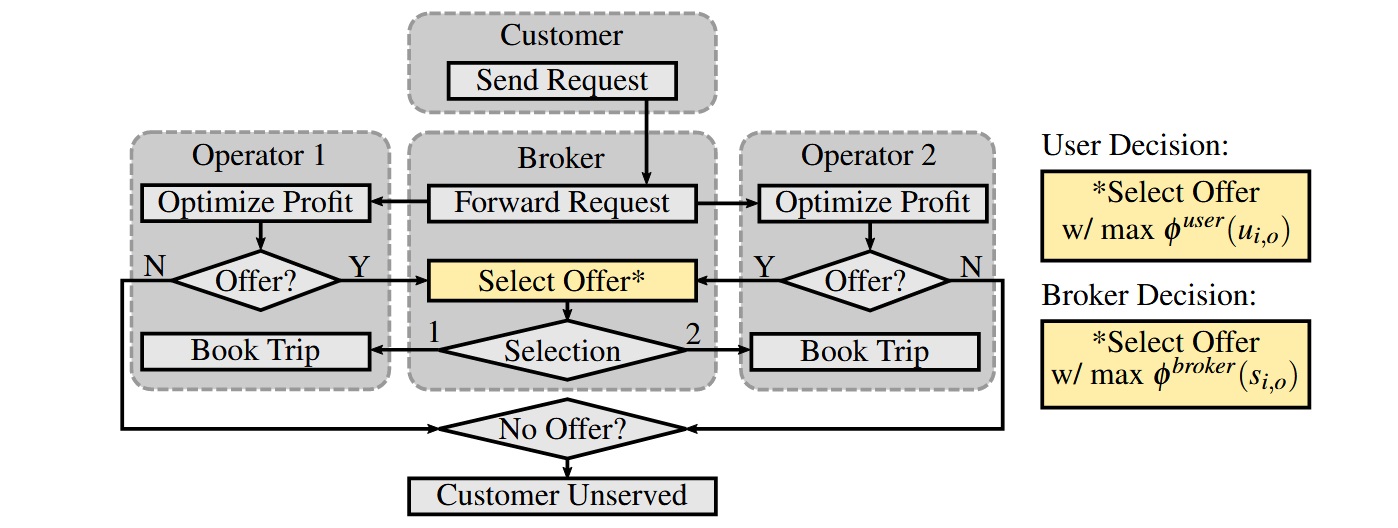}
\end{center}
\caption{Flowchart of the User Decision and Broker Decision scenarios. The only difference between the scenarios is the criteria used to choose an offer, highlighted in yellow.}\label{fig:2}
\end{figure}

\paragraph*{3) \textbf{User Decision:}}
In this scenario, multiple AMOD operators are offering the mobility service over a central platform, here referred to as a ``broker''. Instead of interacting directly with one of the operators, customers request a trip from the broker, which forwards the request to each of the operators. The operators then each send an offer to the broker, which presents these options to the customer. The customer then chooses the offer with the highest user utility $\phi^{user}(u_{i,o})$. If the broker does not receive an offer from either of the operators, the customer leaves the system unserved. A flowchart of this scenario is shown in Figure~\ref{fig:2}.

\paragraph*{4) \textbf{Broker Decision:}}
In this scenario, multiple AMOD operators are also offering the mobility service via a central broker. Customers send their requests to the broker, which then forwards them to each of the operators, who send their offers back to the broker. In contrast to the User Decision scenario, however, rather than allowing the customer to choose their preferred offer, the broker chooses the offer which it deems best for the transportation system by evaluating the highest system utility $\phi^{broker}(s_{i,o})$. Therefore, a broker decision aims towards a system-optimal state, whereas the user decision reflects a quasi-user optimal scenario. A flowchart of this scenario is shown in Figure~\ref{fig:2}.

\subsection{Fleet Operator Model}

The main tasks of each operator are (i) to create offers for customers (or a broker) which serves as their basis to decide for or against the service, (ii) to assign and schedule its vehicles to customers who have booked their service , and (iii) distribute idle vehicles according to expected demand by assigning repositioning tasks.

The assignment of customers to vehicles and their corresponding schedules is modeled as a solution of a dynamic vehicle routing problem. With the set of vehicles $V_o$ of operator $o$, we define a schedule $\psi_k(v, R_\gamma)$ as the $k$-th feasible permutation of stops for vehicle $v \in V_o$ serving the subset of requests $R_\gamma \subset R$ of all currently active requests $R$. Hereby, stops refer to origin and destination locations of requests in $R_\gamma$ where boarding and alighting processes of the corresponding customers are performed. In this study, a schedule is called feasible if
\begin{enumerate}
    \item for each customer, the alighting stop succeeds the boarding stop,
    \item at no point in time the number of on-board passengers exceeds the vehicle capacity $c_v$, 
    \item each customer $i \in R_\gamma$ has been or is scheduled to be picked up before a maximum waiting time $t_{max}^{wait}$ has elapsed after the request time $t_i$, and
    \item the in-vehicle travel time of each customer $i \in R_\gamma$ is not increased by more than $\Delta$ compared to the direct travel time between $x_i^s$ and $x_i^d$.
\end{enumerate}
To compare different schedules, each schedule $\psi_k(v, R_\gamma)$ is rated by an objective function $\rho_\alpha$ which we define in this study by

  \begin{equation}
\label{eq:obj_function}
  \hspace{-1cm}
    \rho_{\alpha} \left(\psi_k(v, R_\gamma) \right) = c_{\alpha}^{dis} \cdot d \left(\psi_k(v, R_\gamma) \right) + c_{\alpha}^{vot} \cdot \left( \sum_{i \in R_\gamma} t_i^{arrival} \left(\psi_k(v, R_\gamma) \right) - t_i \right) - N_R \cdot |R_\gamma| .
    \end{equation}

$d \left(\psi_k(v, R_\gamma) \right)$ is the distance vehicle $v$ has to drive when executing the schedule $\psi_k(v, R_\gamma)$, and $t_i^{arrival} \left(\psi_k(v, R_\gamma) \right)$ is the expected arrival time of customer $i$ according to this schedule. $N_R$ is a large assignment reward to prioritize serving as many customers as possible. $c_{\alpha}^{dis}$ and $c_{\alpha}^{vot}$ are cost factors reflecting the distance cost and the value of time for customers, respectively. The goal of the optimization is to assign schedules to vehicles that minimize the sum of the objective functions of all assigned schedules. Depending on the operational strategy $\alpha$, the weights $c_{\alpha}^{dis}$ and $c_{\alpha}^{vot}$ can be set to prioritize low passenger service times or low fleet mileage. If $c_{\alpha}^{dis} \gg c_{\alpha}^{vot}$ this objective would favor schedules that require only small distances to drive for vehicles. On the other hand, if $c_{\alpha}^{dis} \ll c_{\alpha}^{vot}$, schedules are favored that serve customers as fast as possible.

If all feasible schedules $\psi_k(v, R_\gamma))$ for all vehicles $v$ and all request bundles $R_\gamma$ can be found, an optimization problem can be solved to assign the currently best schedules to the vehicles. By defining a V2RB (vehicle-to-request-bundle) $\Psi(v, R_\gamma)$ as the set of all feasible permutations $k$ of schedules of vehicle $v$ serving $R_\gamma$ with
\begin{equation}
    \rho_{v,\gamma}^{\alpha} = \rho_{\alpha} \left( \Psi(v, R_\gamma) \right) = \min_k \rho_{\alpha} \left(\psi_k(v, R_\gamma) \right) ,
\end{equation}
being the objective function value of this V2RB, an integer linear problem (ILP) can be formulated:
\begin{align}
    \label{eq:opt1}
    \text{minimize} \qquad & \qquad \sum_v \sum_\gamma \rho_{v,\gamma}^{\alpha} \cdot z_{v,\gamma} \\
    \label{eq:opt2}
    \text{s.t.} \qquad & \qquad \sum_\gamma z_{v,\gamma}  \leq 1 \qquad \forall v \in V_o \\
    \label{eq:opt3}
        & \qquad \sum_v \sum_{\gamma \in \Omega_i} z_{v,\gamma}  = 1 \qquad \forall i \in R_a \\
    \label{eq:opt4}
        & \qquad \sum_v \sum_{\gamma \in \Omega_i} z_{v,\gamma}  \leq 1 \qquad \forall i \in R_u \qquad .
\end{align}
Equation~\eqref{eq:opt1} tries to select schedules with cost $\rho_{v,\gamma}^{\alpha}$ of vehicle $v$ to serve the bundle of requests $\gamma$ that minimizes the total cost. Thereby, $z_{v,\gamma} \in \{0,1\}$ is the decision variable taking the value $1$ if schedule with cost $\rho_{v,\gamma}^{\alpha}$ is assigned and $0$ otherwise. Equation~\eqref{eq:opt2} ensures that only one schedule can be assigned to each vehicle. Equation~\eqref{eq:opt3} ensures that each customer $i$ from the set of already assigned customers $R_a$ has to be assigned to exactly one vehicle again. Here, $\Omega_i$ corresponds to all request bundles that include customer $i$. In the same way, Equation~\eqref{eq:opt4} ensures that each customer that has not been assigned yet (set $R_u$) can be assigned to at most one vehicle.

Within the simulation, customers can request trips in every simulation time step of 60 seconds. Depending on the scenario, the customers or the broker decide for or against the service depending on the respective offers sent by the operators. In this study, an immediate decision process is assumed, i.e. the operators are informed that a customer is either booking a trip or declines an offer before the next customer can request a trip. The operator creates offers based on a solution (assigned vehicle schedule including the request) of the optimization problem of Equation~\eqref{eq:opt1}. If no solution can be found, the request is declined by the operator. Since solving the optimization problem for each customer would be computationally intractable, a multi-step approach is applied. In a first step (offer phase), each time a customer requests a trip, a heuristic is applied to find an initial solution for Equation~\eqref{eq:opt1}. This initial solution is used to create an offer. If the customer books the service, the solution (schedule) is assigned to the vehicle, otherwise the solution is discarded. In a second step, after all customer requests in one time step have been processed, a global re-optimization is performed by solving optimization problem~\eqref{eq:opt1} for all currently scheduled or on-board requests. These two steps are described in more detail in the following.

In the offer phase, an insertion heuristic is applied to find the initial solution from which the offer is created. In this heuristic, new feasible vehicle schedules are constructed by inserting customers into the currently assigned vehicle schedules. Because a schedule can only be feasible if this new customer can be picked up within $t_{max}^{wait}$, an insertion need only be tested for vehicles that can reach the customer's origin within this time interval. Let $\psi_k(v, R_\gamma)$ be a feasible insertion of customer $i$ into the current solution of vehicle $v$ and $\psi_l(v, R_{\gamma \setminus i})$ be the current solution of vehicle $v$. The offer is based on the solution of the local optimization problem 
\begin{equation}
  \min_{v, k} \rho_{\alpha} \left(\psi_k(v, R_\gamma) \right) - \rho_{\alpha} \left( \psi_l(v, R_{\gamma \setminus i}) \right) \qquad \forall \text{ feasible }v, k\text{ .}
  \end{equation}

For re-optimizing the vehicle schedules once all new customers within the current simulation step have been processed, an algorithm based on that of \citep{AlonsoMora.2017} is applied in this study. A high level description of the implementation is presented here, while the reader is referred to \cite{Engelhardt.7292020} for details. The idea of the algorithm is to find all feasible schedules first and solve the ILP (Equations~\eqref{eq:opt1}~-~\eqref{eq:opt4}) based on these schedules afterwards. Since an exhaustive search is intractable for the ridepooling assignment problem, a guided search is applied. This guided search can be divided into the following three steps: In a first step, all feasible vehicle-customer combinations are searched. These combinations are defined as feasible for all vehicles theoretically able to reach the origin of the customer request within $t_{max}^{wait}$. In a second step, all feasible customer-customer combinations are searched. A customer-customer combination is defined as feasible if a feasible schedule for a hypothetical vehicle can be found which serves both customers (shared or one-after-the-other). In a third step, the first two algorithm steps are exploited to create all feasible V2RBs (schedules) sorted by their grade, which we define as the number of customers that are served by the corresponding schedules. A V2RB of grade one for vehicle $v$ serving customer $i$ can only exist if the corresponding vehicle-customer combination from the first step is feasible. A V2RB of grade two can only exist if both vehicle-customer combinations between vehicle and customers are feasible and additionally the customer-customer combination is feasible. And finally, a V2RB of grade $n$ can only exist, if all V2RBs of grade $n-1$ exist, where one of the $n$ customers is removed. That is, for a V2RB $\Psi(v, R_\gamma = \{1,2,3\})$ to exist, the feasibility of V2RBs $\Psi(v, R_{\gamma \setminus \{3\}} = \{1,2\})$, $\Psi(v, R_{\gamma \setminus \{1\}} = \{2,3\})$ and $\Psi(v, R_{\gamma \setminus \{2\}} = \{1,3\})$ is necessary. All feasible schedules can now be created iteratively by increasing the grades of the V2RBs.

To adjust the spatial distribution of vehicles for upcoming demand, a repositioning strategy is applied periodically. Every $T^{repo}$, a parameter-free rebalancing strategy based on \citep{Pavone.2012} is applied. After estimating available vehicles and expected demand for each taxi zone, a minimum transportation problem, which aims to minimize the travel costs to reach some zone supply and demand balance constraints, is solved.

\begin{figure}[htbp]
\begin{center}
\includegraphics[width=\textwidth]{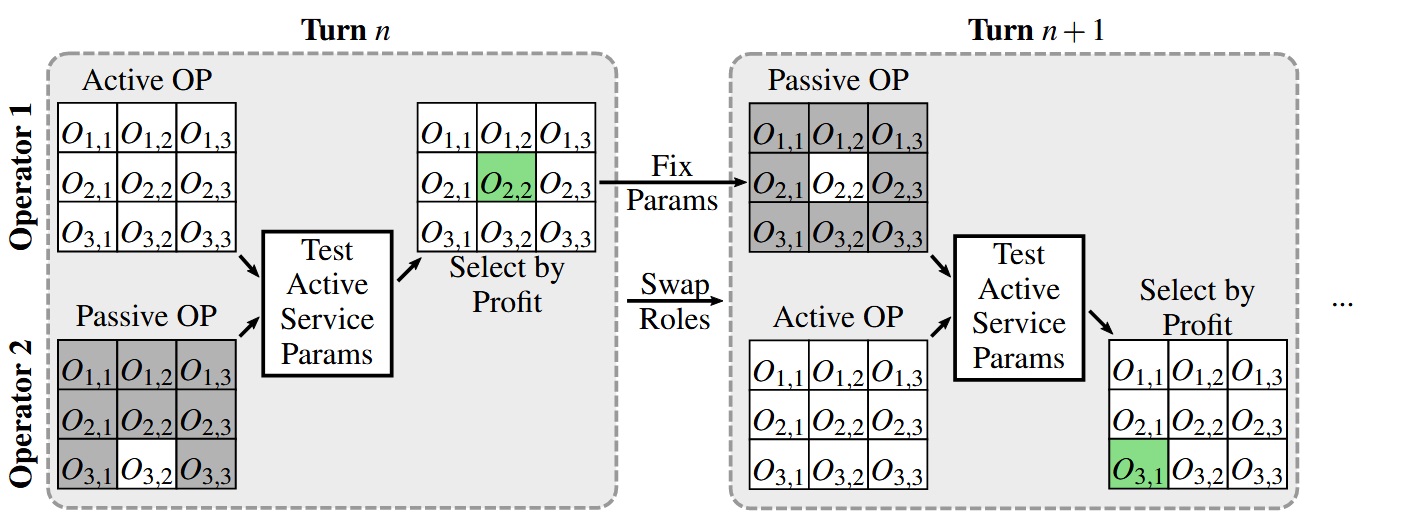}
\end{center}
\caption{Illustration of the game. Operators take turns playing the ``active'' role. In each turn, the active operator tests all of its possible service parameters $O_{m,n}$ against the passive operator's fixed parameters. The parameters that bring the highest effective profit for the active operator are then used in the next turn, where the roles are reversed.}\label{fig:3}
\end{figure}

\subsection{Game}

The different interaction scenarios introduced in the beginning of this section describe different external environments in which the operators offer their service. Depending on these environments operators will adapt their service design to maximize profit. In this study, the adaptation of the service of the operators is modeled as a turn-based game (illustrated in Figure~\ref{fig:3}). Each operator starts with specific service parameters based on the scenario without interaction. In each turn, one operator has the active role, while the other one has the passive role. These roles are exchanged every turn. The active operator explores different sets of service parameters (with exhaustive search), while the service parameters of the passive operator remain constant. At the end of each turn, the active operator adopts the service parameters that resulted in the highest profit.

The profit $P$ is calculated by the difference of revenue $R$ and costs $C$ after each simulation.
\begin{align}
  \label{eq:profit}
    P &= R - C \\
    R &= \sum_{i \in C_{served}} d^{direct}_i \cdot f \\
    C &= N_v \cdot C_v + d^{fleet} \cdot c_{dis},
 \end{align}
with $C_{served}$ being the set of all served customers, $d^{direct}_i$ their corresponding direct travel distances, and $f$ a distance-dependent fare the customers have paid. $C_v$ is the fixed cost per vehicle, $d^{fleet}$ is the driven distance of the vehicle fleet and $c_{dis}$ is the distance-dependent vehicle cost.

To optimize this profit $P$ for only a few days of simulation, operators would choose small fleets to increase overall vehicle utilization (including off-peak times). However, the service rate would suffer strongly leading to an unreasonable large number of customers that have to be rejected. Since such an unreliable service is improbable to survive on the long run, another term penalizing the number of requests, which did not receive an offer $N_{C, no}$ (within the given service quality constraints) during the simulation, should be considered. Hence, we define the effective profit $P_{eff}$ to be maximized within the game:
\begin{equation}
  \label{eq:effprofit}
    P_{eff} = P - N_{C, no} \cdot p_{no} .
 \end{equation}
Thereby, $p_{no}$ is a penalty cost for each request without offer. $p_{no}$ and $f$ will be determined within calibration simulations in the results section.

Alternating turns with operators maximizing their effective profit are repeated until equilibrium is reached. All operators adopting the same parameter set is one of the equilibrium states the game can converge in. In this case no operator has an advantage in changing their parameter sets anymore. Generally, it is not guaranteed to find such equilibrium states. Depending on the overall setting, it is for example feasible that operators with different market penetrations and therefore different fleet sizes also reflect a stable system. Nevertheless, it turns out that as long as certain symmetries between the operators are assumed as described in the following case study, these symmetric equilibrium states can be found in all scenarios tested.

\section{Case Study}

We test the model on a case study for the publicly available taxi data set of Manhattan, NYC. The simulation inputs are described in the following.

\begin{figure}[htbp]
\begin{center}
\includegraphics[width=\textwidth]{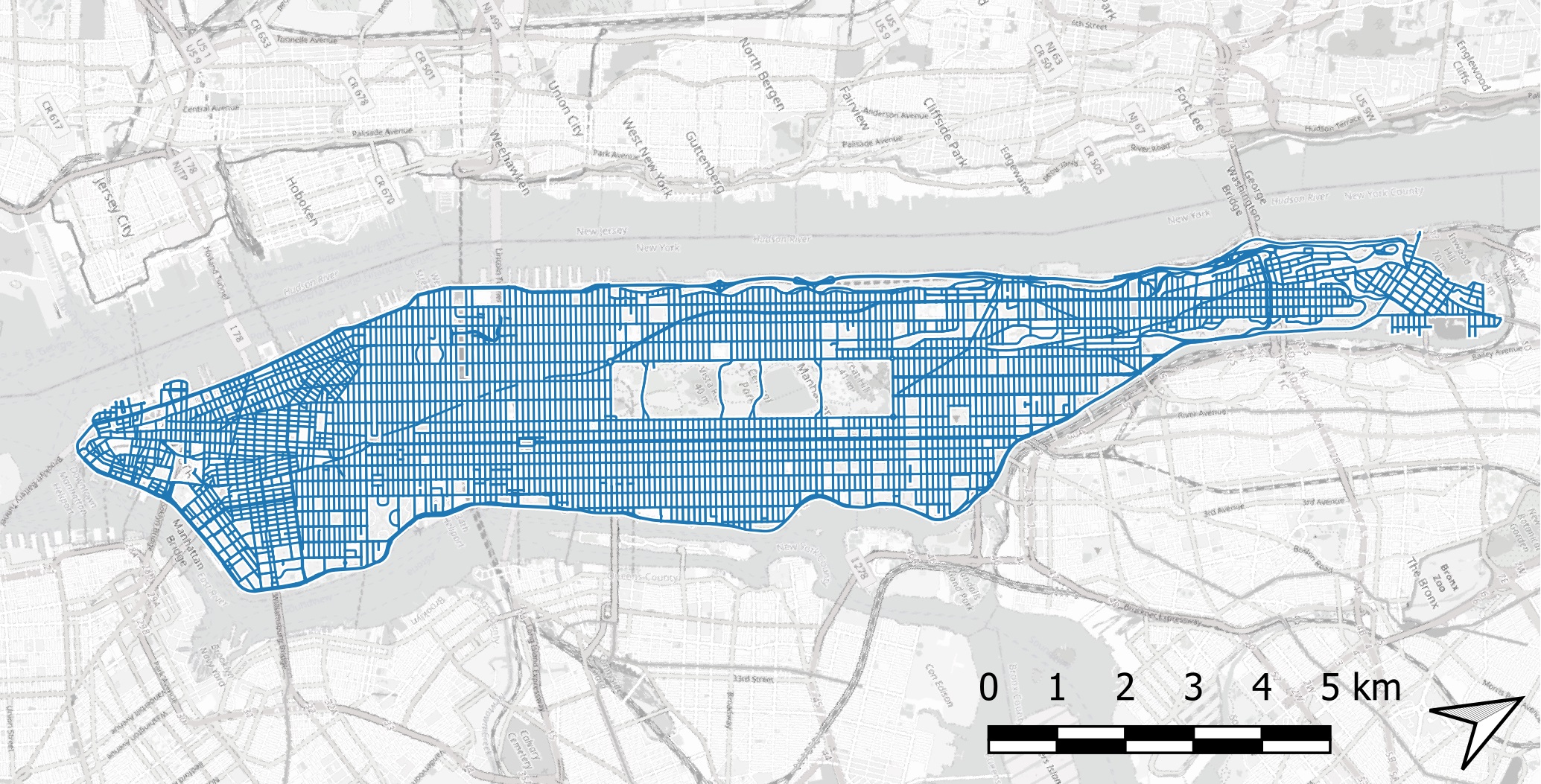}
\end{center}
\caption{Street network of Manhattan, NYC used in the case study.}
\label{fig:4}
\end{figure}

\paragraph*{\textbf{Network and Demand:}}
Figure~\ref{fig:4} shows the street network and the operating area of the simulated ridepooling services. All operators offer the service in the same operating area in this study. The street network $G=(N,E)$ has been extracted from OpenStreetMap data using the Python library OSMnx \cite{Boeing.2017}. Initially, edge travel times are assigned according to free flow speeds from the OpenStreetMap data. In order to replicate realistic travel times, edge travel times are scaled after every 15 min simulation time according to actual trip travel times within the NYC taxi trip data. Shortest (travel time) paths are computed using a combination of the classical Dijkstra algorithm and preprocessing origin-destination node pairs in lookup tables.

As demand for the ride pooling service, NYC taxi trips that are starting and ending within the operating area of Manhattan are used. Trip requests are created for the week from 2018/11/11 to 2018/11/18. Trip origins and destinations are matched onto the closest intersection nodes that are only connected to roads with classes “living street”, “residential”, “primary”, “secondary”, and “tertiary”. Presumably defective trip records with average travel times below $1$~m/s or above $30$~m/s are removed from the data set. Overall 1511476 trips remain in the data set. To decrease overall computational time, this set is subsampled to generate the requests for the ridepooling services: For each trip a random number between $[0, 1[$ is drawn. If this random number is smaller than $0.1$, the trip is transferred into the set of ridepooling requests resembling a $10$\% market penetration of the simulated ridepooling services. Using different random seeds, three set of request sets are generated and used within the simulations.

The rebalancing algorithm is called every $T^{repo} = 15$~min. Demand and supply forecasts are aggregated to the corresponding taxi zones. For simplicity, trip forecasts, i.e. the average number of incoming and outgoing trips within a time interval of $15$~min per zone, are created by counting the overall trips in the data and multiplying the counts with the market penetration of $10$\%. In the case of multiple operators sharing the demand, it is assumed that all operators rebalance the vehicle fleet based on the same spatio-temporal forecast distribution. Therefore the average counts are additionally divided by the number of operators.

Further details on network and trip data processing can be found in \citep{Syed.2021}.

\paragraph*{\textbf{Scenario Specification:}}
We evaluate a system with a maximum of two ridepooling operators. It is assumed in this study that both operators offer a similar service quality. Namely, the operators employ vehicles with maximum traveler capacity $c_v = 4$. Additionally, they only offer trips to customers that do not exceed (i) a maximum waiting time of $t_{wait}^{max} = 6$~min, and (ii) a relative increase in travel time $\Delta = 40$\% compared to the duration of a direct trip. Because a similar service is offered by the two ridepooling providers, we additionally assume that: (i) customers do not have an inherent preference towards any particular operator, and (ii) due to market pressure, both operators synchronize their fares and offer their service for the same prices. Therefore, price sensitivity of customers is not explicitly modeled.

With respect to the different interaction scenarios, the inputs for these scenarios are the following:
\begin{enumerate}
    \item Single Operator: A single operator with the specified attributes serves the whole demand.
    \item Independent Operators: The demand is split evenly between two operators. Each customer can only request a trip from the corresponding assigned operator.
    \item User Decision: The broker forwards customer requests to both operators. In case a customer $i$ receives offers from both operators, the decision to book with operator $o_i$ is made based on the evaluation of
    \begin{equation}
        o_i = \argmax_{o} \phi^{user}(u_{i,o}) = \argmin_{o} t^{arr}_{i,o},
     \end{equation}
    with the arrival time $t^{arr}_{i,o}$ offered by operator $o$.
    \item Broker Decision: The broker requests trips for the customers from each operator. In this study, the system costs are measured by the additional driven distance to accommodate a new request. Hence, in case the broker receives offers from both operators, the decision to book customer $i$ with operator $o_i$ is made based on the evaluation of
    \begin{equation}
        o_i = \argmax_{o} \phi^{broker}(s_{i,o}) = \argmin_{o} \delta d_{i, o},
        \end{equation}
    with the additional driving distance $\delta d_{i, o}$ required for operator $o$ to serve customer $i$.
\end{enumerate}

The parameters defining the objective function for each operator are set to $c_{\alpha}^{dis} = 0.25$\euro/km and $c_{\alpha}^{vot} = 16.2$\euro/h, corresponding to the estimated values in \cite{Bosch.2018} and \cite{Frei.2017}, respectively.

\paragraph*{\textbf{Game:}}

The goal of the game is to model operators' adaptation of their service within different environments (interaction scenarios) to maximize their profit. While there are many different parameters for operators to adapt, in this study we allow the operators (i) to change their fleet size and (ii) to modify their objective function for assigning offers and vehicle plans. Fleet sizes $N_v$ can be changed initially in steps of 20 vehicles around the initial fleet size to be defined in the following calibration step. In the ``Single Operator'' scenario, one operator has to serve double the amount of requests; hence, fleet size step sizes are doubled accordingly. Possible parameter options $(c_{\alpha}^{dis}, c_{\alpha}^{vot})$ for setting the objective function from Equation~\eqref{eq:obj_function} are ($0.0$\euro~/km, $16.2$\euro/h), ($12.5$\euro/km, $16.2$\euro/h), ($25.0$\euro/km, $16,2$\euro/h), ($25.0$\euro/km, $8.1$\euro/h), and ($25.0$\euro/km, $0.0$\euro/h) initially. With these options, the objective function can be adapted quite smoothly between purely minimizing the driven distance to purely minimizing customer arrival times.

Once an equilibrium with the initial parameter step sizes can be observed, the step sizes for fleet size and objective parameters are decreased for the remaining steps of the game to increase the resolution quality of the equilibrium state. Thereby, parameter steps are adopted by halving the step size, setting the currently found optimum within the new parameter table to be observed. This procedure is repeated until no clear symmetric equilibrium can be found anymore, which is interpreted as the maximum solution quality possible with respect to stochastic variations within the simulations. In the conducted simulations, alternating jumps between neighbouring cells in the parameter table are observed indicating the best possible resolution quality of the optimal parameter set.

Parameters for calculating the costs in Equation~\eqref{eq:profit}, i.e. the fix cost per vehicle $C_v$ and the distance-dependent cost $c_{dis}$, are set to $25$\euro per day and $c_{dis} = 25.0$\euro/km, respectively, according to \cite{Bosch.2018}.

The fare $f$ to calculate the profit in Equation~\eqref{eq:profit} and the penalty cost for requests without offer $p_{no}$, which are required to calculate the effective profit in Equation~\eqref{eq:effprofit}, will be determined within calibration simulations in the next chapter.

\section{Results}

In this section, results of the simulations are presented. Firstly, the calibration is described to determine the initial fleet size as well as the parameters $p_{no}$ and $f$. Secondly, the results after performing the game are presented and lastly, fleet key performance indicators (KPIs) are compared before and after the game and between the different interaction scenarios.

\subsection{Calibration}

\begin{figure}[htbp]
\begin{center}
\includegraphics[width=\textwidth]{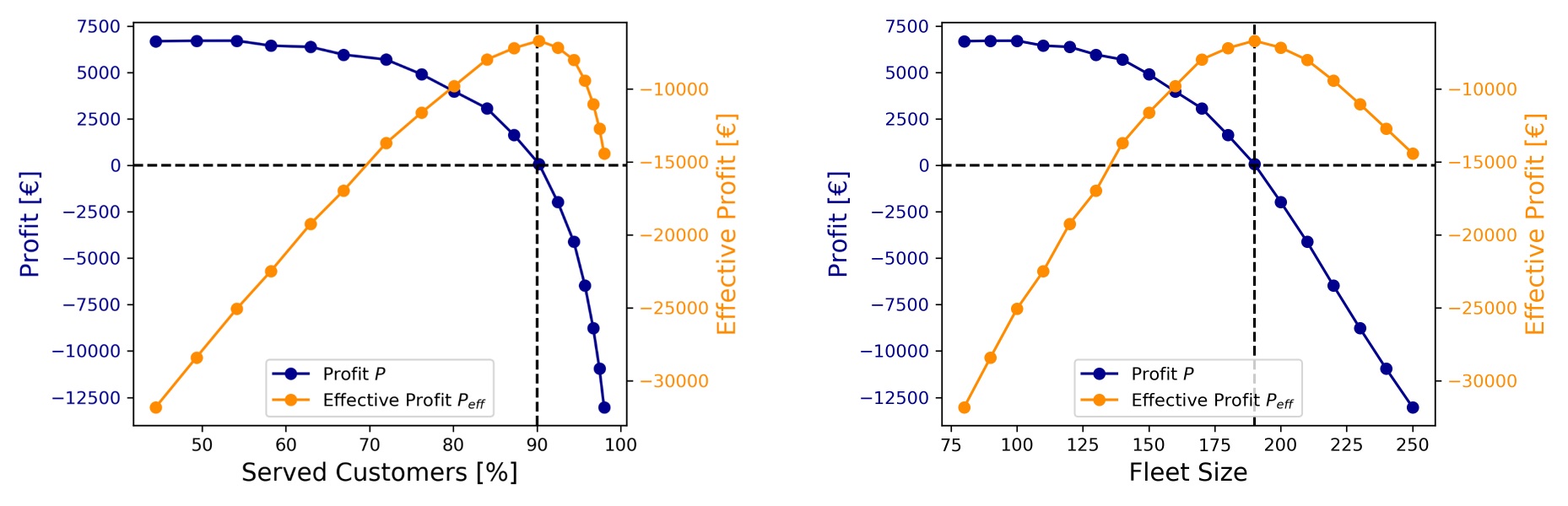}
\end{center}
\caption{Calibration of break even fare and unserved customer penalty. The break even fare is chosen to achieve $0$~\euro ~ Profit at $90$\% served customers, while the unserved customer penalty is set to result in a maximum for the Effective Profit at $90$\% served customers as shown in the left figure. $190$ vehicles are needed for each operator to served $90$\% customers as depicted in the right figure.}
\label{fig:5}
\end{figure}

Since the envisioned autonomous ridepooling services are not yet operating, the values for $f$ and $p_{no}$ cannot be found empirically. Instead, we use the interaction scenario of two independent operators as calibration scenario, where we choose $90$\% served customers as a target service rate. Conducting simulations for fleet sizes ranging from 75 to 250 reveals 190 vehicles are needed for each operator to achieve this service rate. The distance dependent fare $f$ is chosen to create a break even profit using 190 vehicles resulting in $f = 43$~ct/km (see blue curve in Figure~\ref{fig:5}). The goal of calibrating the penalty cost parameter $p_{no}$ for unserved requests is to create a maximum for effective profit $P_{eff}$ at the target service rate of $90$\%. A value of $p_{no} = 46$~ct accomplishes this target (see orange curve in Figure~\ref{fig:5}) and is used for further simulations.

\subsection{Game}

\begin{figure}[h!tbp]
\begin{center}
\includegraphics[width=0.85\textwidth]{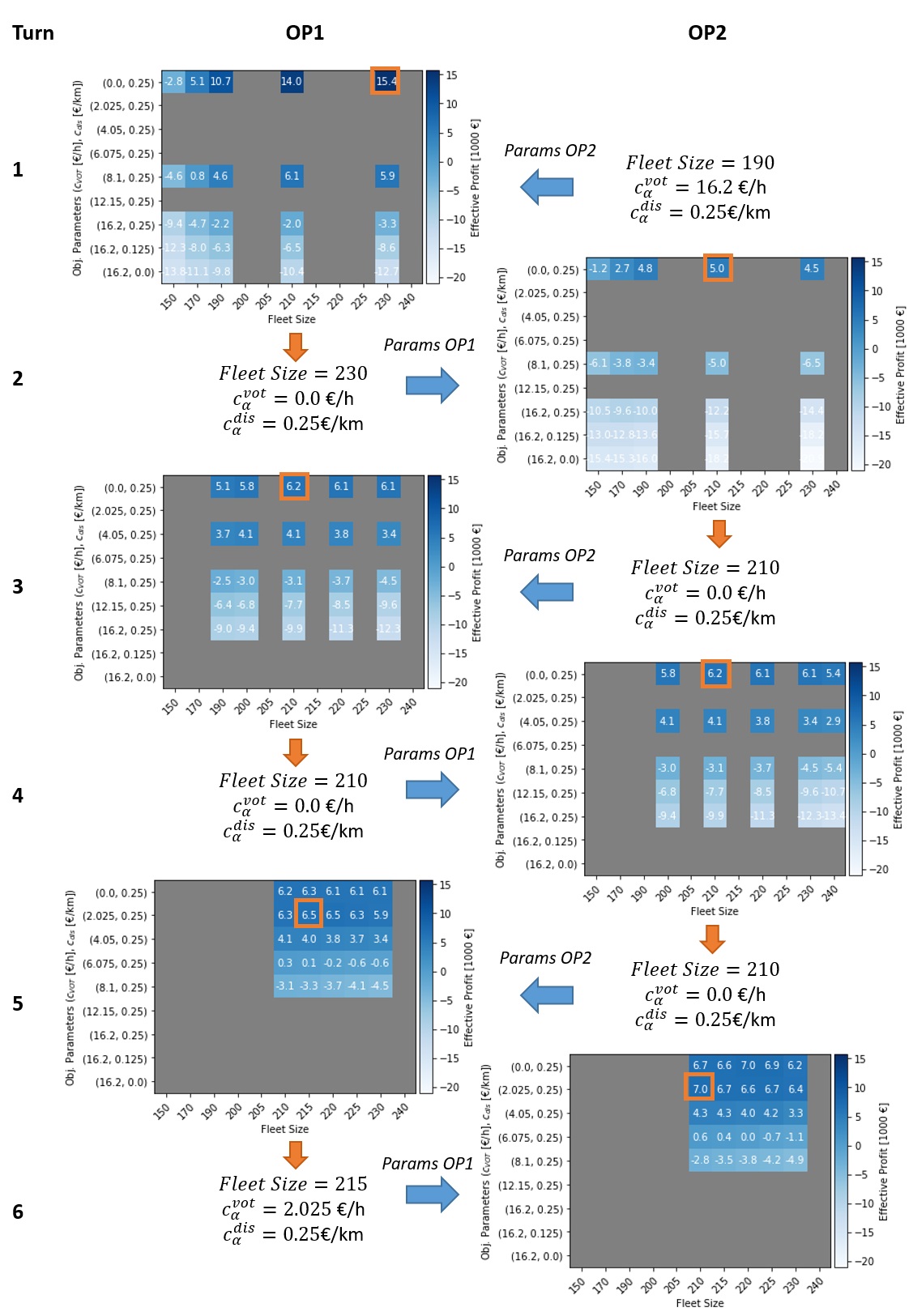}
\end{center}
\caption{Illustration of the development of operator parameters over the course of the game for the broker interaction scenario. Blue arrows indicate parameter settings of the passive operator. Orange arrows indicate the parameter selection resulting in the highest effective profit of the active operator. A first equilibrium can be observed at turn 4, when OP2 adjusts its parameters to the same parameters as OP1. After turn 6 alternating behavior is revealed once the step size is decreased further in turn 5.}
\label{fig:6}
\end{figure}

Figure~\ref{fig:6} shows the development of operator service parameters over the course of the game for the broker scenario. Within each turn, the active operator explores 6 by 6 different possibilities for fleet size and objective function parameters, respectively, while the parameters of the passive operator remain fixed. During the course of the game the differences between neighboring explored parameter possibilities (in the region of the optimum in the rougher grid) become smaller to increase accuracy. This is illustrated by grey fields in Figure~\ref{fig:6} as yet unexplored combinations. After each turn, the active operator takes over the parameter set resulting in the highest effective profit indicated by the orange boxes. For all interaction scenarios an equilibrium can be observed by no later than 6 turns. As indicated in Figure~\ref{fig:6} no clear symmetric equilibrium is observed after increasing the step accuracy, which is also the case for the user decision interaction scenario. Instead alternating jumps between neighbouring cells are observed, which can likely be attributed to the dynamic and stochastic nature of the agent-based simulation model. In the shown example of the broker scenario, simulations until turn 10 reveal alternating jumps within the cells (Fleet Size = $210$ veh, $c_{\alpha}^{vot} = 2.025$~\euro/h, $c_{\alpha}^{dis} = 0.25$~\euro/km) and (Fleet Size = $215$ veh, $c_{\alpha}^{vot} = 2.025$~\euro/h, $c_{\alpha}^{dis} = 0.25$~\euro/km). For reasons of symmetry and because only jumps between neighboring cells occur, symmetric operator parameters are assumed for further evaluation. The parameter set after the first jump to neighbouring cells is applied for both operators. In the case of the example in Figure~\ref{fig:6}, this leads to a final parameter set of (Fleet Size = $210$ veh, $c_{\alpha}^{vot} = 2.025$~\euro/h, $c_{\alpha}^{dis} = 0.25$~\euro/km) in turn 6.

Table~\ref{tab:1} shows operator parameters before and after the game. Initially all operators start with a fleet size of 190 vehicles, or 380 vehicles in the case of a single operator, and an objective function parameterized by $c_{\alpha}^{vot} = 16.2$~\euro/h, $c_{\alpha}^{dis} = 0.25$~\euro/km. For a single operator, the game breaks down to a single round with one optimization table for each parameter set to be explored. As a result of the optimization, the single monopolistic operator decreases the fleet size and the weight of $c_{\alpha}^{vot}$. Due to scaling effects of ridepooling services the larger demand can be served more efficiently, and the cost reduction of operating a smaller fleet (viewed relatively) thereby exceeds the overall penalty of unserved customers. Additionally, costs for driven distance can be reduced without direct competition by decreasing the weight of $c_{\alpha}^{vot}$, thereby putting more focus on decreasing fleet mileage and increased pooling efficiency rather than fast customer pickup and delivery. Similar behavior for adjusting the objective function can be observed for two independent operators. Nevertheless, they even have to slightly increase their fleet size because the fleet can be used less efficient when demand is shared between the operators. Within the user decision scenario, operators are in direct competition against each other. For a customer to book a ride with a specific operator, the operator has to offer the smallest combined waiting and travel time. Therefore, operators select the highest value for $c_{\alpha}^{vot}$ in this scenario to assign routes with small customer arrival time. In the broker decision scenario, operators are also in direct competition to each other, but the decision for a customer to book with one of the operators is based on the offer with the smallest additional driven distance. Similarly to the scenario with independent operators, the value for $c_{\alpha}^{vot}$ is decreased which puts a higher relative weight on the distance cost factor $c_{\alpha}^{dis}$ for assigning routes. Compared to the other interaction scenarios, operators tend to have the highest fleet sizes in the case of the broker decision. A higher density of vehicles will lead to shorter pick-up trips (on average) and seems preferable in this scenario.

\begin{table}[]
\begin{center}
\begin{tabularx}{.92\textwidth}{lllllll}
    \toprule
                   & \multicolumn{2}{c}{Single Operator Scenario}                          & \multicolumn{4}{c}{Multi-Operator Scenario}                                                                                                                    \\
                   \cmidrule(lr){2-3} \cmidrule(lr){4-7}
                   & \multicolumn{1}{c|}{Initial} & \multicolumn{1}{c}{Final}              & \multicolumn{1}{c|}{Initial} & \multicolumn{3}{c}{Final}                                                                                                        \\
\textbf{Parameter} & \multicolumn{1}{c|}{}        & \multicolumn{1}{c}{\textbf{Single Op}} & \multicolumn{1}{c|}{}        & \multicolumn{1}{c}{\textbf{Indep. Ops}} & \multicolumn{1}{c}{\textbf{User Dec.}} & \multicolumn{1}{c}{\textbf{Broker Dec.}} \\
\midrule
Fleet Size [veh]         & 380                          &     310                                    & 190                          &    195                                          &     195                                   &  210                                        \\
$c_{\alpha}^{vot}$ [\euro/h]          & 16.2                         &      2.025                                   & 16.2                        &    2.025                                          &     6.075                                   &    2.025                                      \\
$c_{\alpha}^{dis}$ [\euro/km]          & 0.25                        &      0.25                                   & 0.25                        &     0.25                                         &      0.25                                  &    0.25    \\                                 
\bottomrule
\end{tabularx}
\caption{Operator service parameters before (Initial) and after the game (Final).}
\label{tab:1}
\end{center}
\end{table}

\subsection{Fleet KPIs}

\begin{figure}[htbp]
\begin{center}
\includegraphics[width=0.8\textwidth]{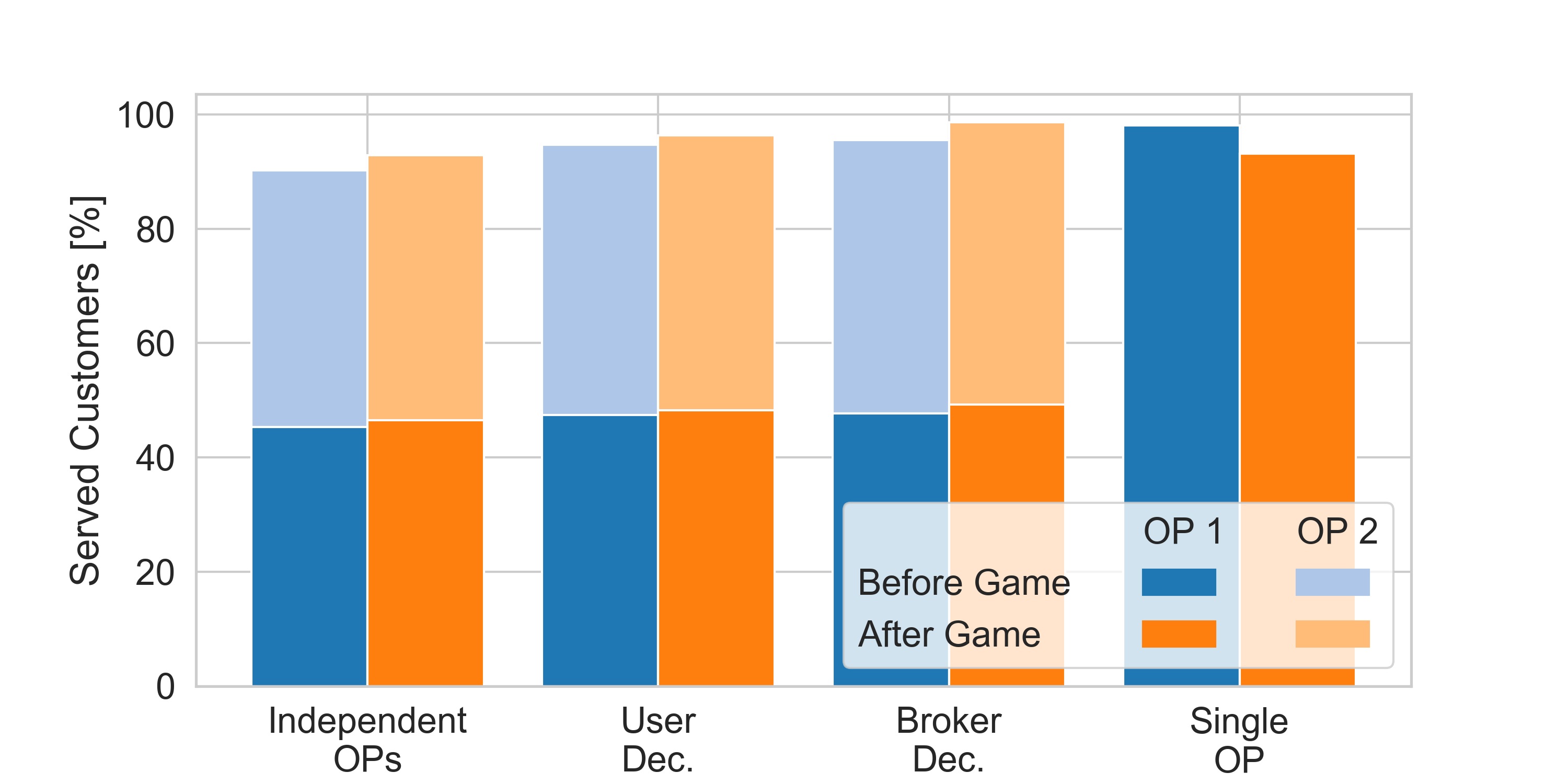}
\end{center}
\caption{Served Customers before and after the game for the different interaction scenarios.}
\label{fig:7}
\end{figure}

Figure~\ref{fig:7} shows the fraction of served requests before and after the game for each interaction scenario. In all cases around $90$\% of all requests could be served as targeted within the calibration. Before the game the overall fleet size in the system is set the same for all interaction scenarios to illustrate the price of non-coordination. Therefore, most customers could be served within a single monopolistic operator setting, because the fleet can be controlled most efficiently having full access to all customers. On the contrary, with completely independent operators fewest customers can be served before the game due to effects of market fragmentation. Because customers have access to both operators and can choose the other operator in case the first cannot serve them, in the broker and user decision scenarios the fraction of served customers lies in between. After the game, the single operator decreases its fleet size resulting in the lowest fraction of served customers. Most customers are served in the broker decision scenario because operating larger vehicle fleets is profitable in this case indicating an advantage also for customers in this regulated scenario.

\begin{figure}[htbp]
\begin{center}
\includegraphics[width=0.8\textwidth]{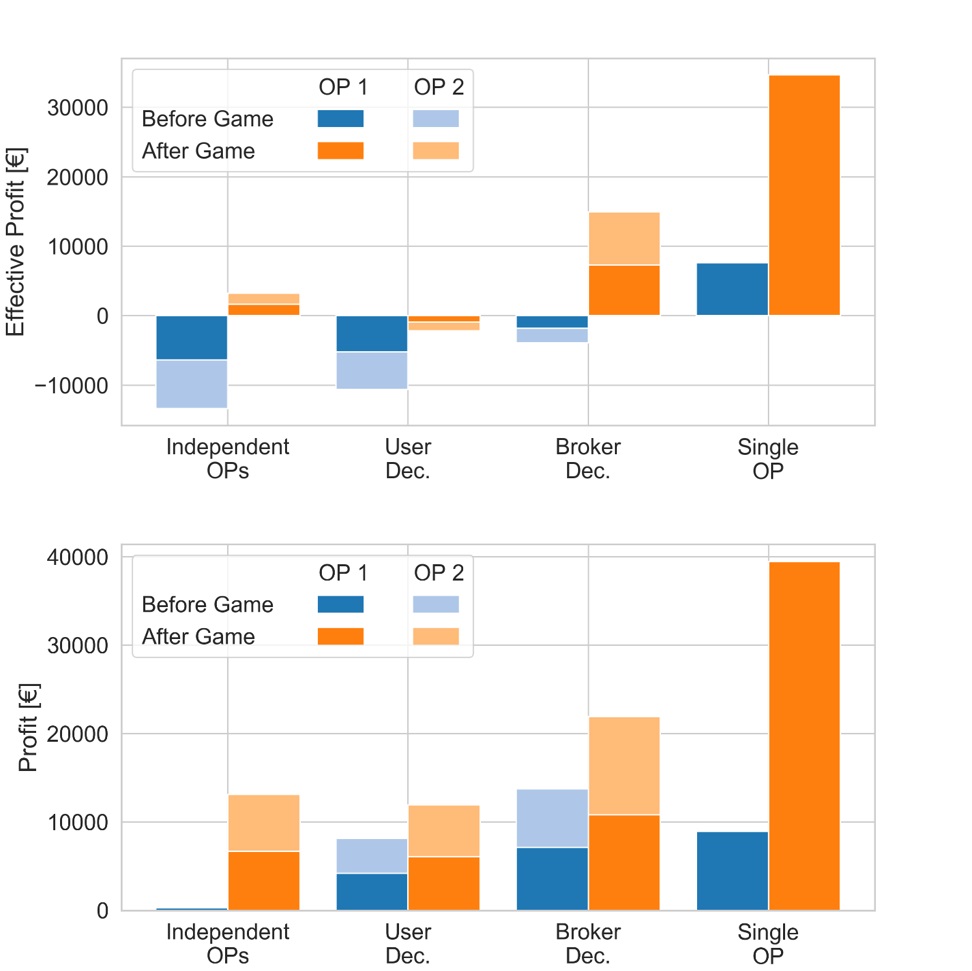}
\end{center}
\caption{Effective Profit (top) and Profit (bottom) before and after the game for the different interaction scenarios.}
\label{fig:8}
\end{figure}

In Figure~\ref{fig:8} the effective profit and the actual profit before and after the game is illustrated. Before the game the effective profit is dominated by the penalty for unserved customers resulting --- similar to Figure~\ref{fig:7} --- in the highest value for the single operator and the lowest one for independent operators. The highest combined actual profit can be obtained within the broker decision scenario. The operator assignment process of selecting the operator with the lowest additional driven distance is here in line with the distance-dependent operating cost. The profit for independent operators is close to zero because this scenario is chosen in the calibration step to define the break even fare. 

After the game, operators could increase their effective as well as their actual profit in all interaction scenarios. The highest gain is obtained for the single operator who could decrease both fixed costs by decreasing fleet size and distance-dependent costs by changing the objective to select shorter routes without losing many customers by competition. The smallest gain is observed in the user decision case. Because of pressure due to competition, operators have to focus on assigning routes with low waiting and detour time for customers which results in in a trade-off to higher fleet mileage and therefore in higher costs. Within all scenarios with more than one operator, operators achieve most actual as well as effective profit in the broker decision setting after the game. On the one hand, assigning customers to operators with the smallest additional driven distance is equal to the option that produces the lowest costs for the operator. On the other hand operators can additionally change their objective to putting more focus on assigning short routes without the market pressure from customers deciding for fastest trips.

\begin{figure}[htbp]
\begin{center}
\includegraphics[width=0.8\textwidth]{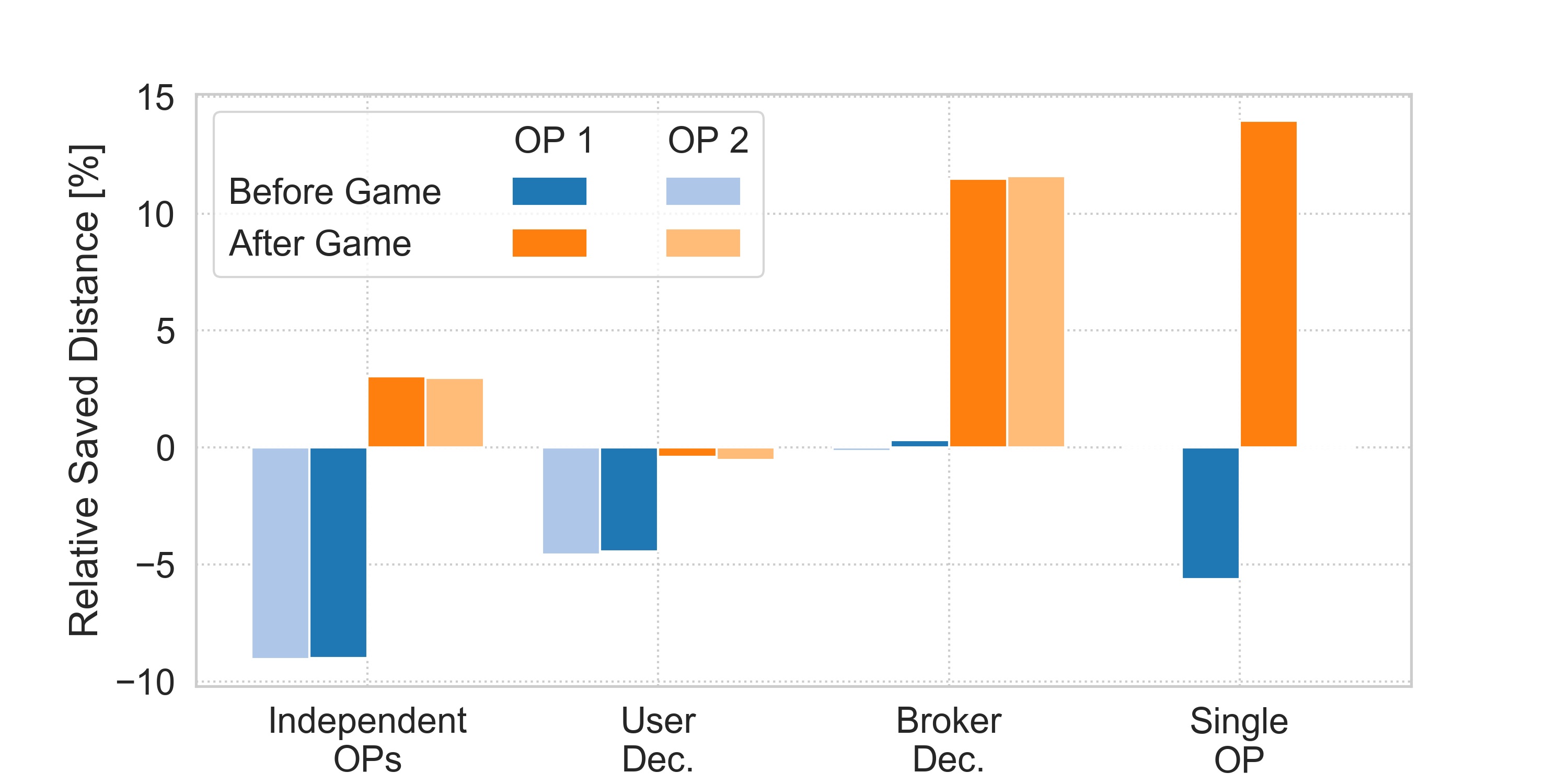}
\end{center}
\caption{Saved Distance before and after the game for the different interaction scenarios.}
\label{fig:9}
\end{figure}

The effectiveness of pooling can be measured by the relative saved distance $rsd$, which is plotted in Figure~\ref{fig:9} and defined by
\begin{equation}
    rsd = \frac{\sum_{i \in C_{served}} d^{direct}_i - d^{fleet}}{\sum_{i \in C_{served}} d^{direct}_i} , 
    \end{equation}
with the direct distance $d^{direct}_i$ of each served customer $C_{served}$ and the fleet driven distance $d^{fleet}$. The higher this quantity is, the higher the fraction of fleet driven distance that has been shared between customers. However, in contrast to simply evaluating the average occupancy, unreasonable detours with multiple passengers on board do not improve this performance indicator. Before the game the saved distance of all operators is below or close to zero for all interaction scenarios indicating that the fleet would actually drive more distance than if customers would drive on a direct trip on their own. The main reason is that before the game the objective weight of $c_{\alpha}^{vot}$ is very high compared to after the game resulting in the preference towards direct trips contrarily to pooled trips. After the game the relative saved distance could be improved for all interaction scenarios mainly because all operators decreased their objective weight $c_{\alpha}^{vot}$. The highest value for $rsd$ is measured for the single operator scenario where most pooling can be realized with a centralized option for optimization. The lowest value is observed in the case of user decisions. Here, the operators are forced to keep a rather high value for $c_{\alpha}^{vot}$. Additionally, if multiple options for a trip are available, customers tend to choose trips without pooling because these trips would in many cases result in longer waiting and detour times. The pooling efficiency in the broker decision scenario is nearly as high as with a single operator. With a combination of operators adjusting their parameters accordingly (low value of $c_{\alpha}^{vot}$ and higher fleet size) and the broker preferring pooled ride options, the pooling efficiency lost due to market fragmentation can nearly be restored.

\begin{figure}[htbp]
\begin{center}
\includegraphics[width=0.8\textwidth]{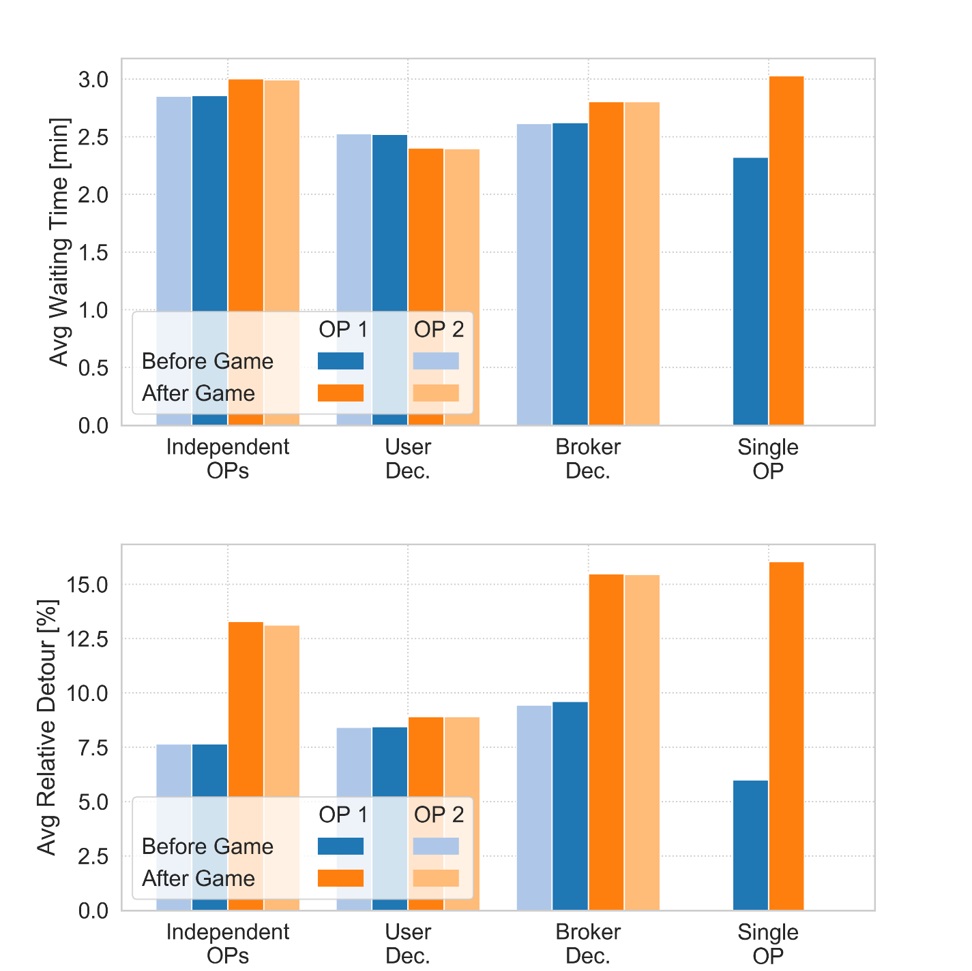}
\end{center}
\caption{Customer waiting time (top) and relative detour time (bottom) before and after the game for the different interaction scenarios.}
\label{fig:10}
\end{figure}

Lastly, Figure~\ref{fig:10} shows customer waiting and detour times. Before the game the average relative detour per customer is rather low indicating few pooled trips, in line with the evaluation of the relative saved distance of Figure~\ref{fig:9}. While the change in customer waiting times comparing before and after the game are minor in all scenarios, a large increase in detour times can be observed especially in the single operator and broker decision scenario. In these scenarios also the relative saved distance increases most, showing the trade-off between customer travel time and efficiency of sharing rides. Nevertheless, the average relative detour of up to $15$\% is still acceptable as it is limited by constraints to $40$\%. Comparing the scenarios after the game, the lowest customer waiting and detour times can be observed for the user decision scenario. Here, customers pick offers with the smallest waiting and travel times while operators additionally put more focus on assigning routes that minimize these parameters.

\section{Conclusion}
\subsection{Summary and Implications}
This study evaluates and quantifies the negative operational impacts of fragmenting AMOD ridepooling demand. Moreover, the concept of an AMOD broker is introduced to counteract these negative aspects. Two extreme forms of this broker, in which (i) the broker only collects the offers of multiple operators and the user selects the operator, and (ii) the broker selects the operator better suited from a system viewpoint, are evaluated in a case study for taxi trips in Manhattan, NYC. The evaluation is performed with agent-based simulations in a static setting with constant fleet size and operator parameters, as well as in a game setting allowing the operators to adapt their service to maximize profit. 

After operators adopted their service parameters in the game setting, the cumulative AMOD fleet size increased to 390 vehicles in the user decision scenario and 420 vehicles in the broker decision scenarios, compared to 310 vehicles in the single-operator system. These increased fleet sizes correspondingly resulted in higher service-rates when competition is present. In most interaction scenarios, operators increase their weight on minimizing fleet mileage to save costs. Only in the scenario where customers choose their AMOD service, operators are forced to offer trips with fast customer arrival times to succeed in competition. Correspondingly, in the user decision scenario pooling efficiency measured by the relative saved distance is reduced by around 14\% compared to a single-operator system. This result indicates, that operators might prioritize offering non-shared trips when competition based on customer decision is present. Contrarily, the broker successfully shifts the operators objectives to decrease fleet mileage resulting in only 2\% in relative saved distance loss compared to the single-operator system. Due to the correlation of fleet mileage and operating costs, most profit could be made when the broker regulates the competition of multiple operators.

All in all, it can be observed that regulating the broker platform can be beneficial for everyone: the transportation system has to supply fewer driven kilometers, the operators can save operational costs, thereby offsetting the fixed costs of adding more vehicles to the fleet, and the users experience a better level-of-service from the additional AMOD vehicles.

\subsection{Future Work}
Several open questions have to be addressed before a real-world application makes sense:
\begin{enumerate}
    \item Who should operate a broker platform?
    \item How will users respond to a platform making the decision for them about which AMOD provider serves them?
    \item Are the shown financial benefits enough motivation for AMOD providers to join a broker, or will municipalities have to enforce it?
\end{enumerate}
The components of the broker objective could be more sophisticated than just additional driven distance, and additionally have to be traceable, i.e. allow a live tracking of fleet KPIs. The complexity of operating such platform is rather high and might be challenging for municipalities. However, as the broker objective and regulations should be aligned with public goals, a private platform provider at least should be paid for and commissioned by municipalities. To avoid cherry-picking, the AMOD services should likely receive a certain level-of-service / share of served request goal, e.g. by adding penalties for requests for which they make no offer. Moreover, an integration into existing public transportation services will be studied in the future. 

With respect to the second question, behavioral studies have to be employed. In addition to the extreme regulatory measure of the broker choosing the AMOD operator, some intermediary levels of regulation can be studied in future work, in which the broker does not impose the choice of operator, but rather merely manipulates the offers (e.g. fares) to influence traveler behavior. Within this context, the symmetry between AMOD service levels should also be relaxed. That is, a broker should also be able to make valuable decisions in case one operator offers a service with high customer convenience, higher fares and lower occupancy compared to a service with the opposite strategy. To quantify such systems, more advanced mode choice models will be required. 

When demand is modeled as price-sensitive, it also makes sense to integrate competitive pricing into the game framework. Moreover, the effect of different repositioning strategies can affect results significantly. It will be interesting to evaluate whether users benefit from competition as trade-offs are likely: users likely experience cheaper fares from competitive pricing, but might also suffer from lower service quality due to reduced ridepooling efficiency resulting from market fragmentation. Moreover, future work could also include studies with more than 2 operators. Asymmetric service design or even strongly asymmetric initial conditions can hint at whether the AMOD market will steer towards monopolies or a shared market with broker platforms. Nevertheless, the applied game framework might not be suited to investigate asymmetric final operator states. Therefore, the application limits of the current game setting have to be studied, and possibly, new methods have to be developed to study these effects.

\section*{Conflict of Interest Statement}

The authors declare that the research was conducted in the absence of any commercial or financial relationships that could be construed as a potential conflict of interest.

\section*{Author Contributions}
Study conception and design: RE, PM, FD, KB; data collection: RE, PM; analysis and interpretation of results: RE, PM, FD; draft manuscript preparation: RE, PM, FD. All authors reviewed the results and approved the final version of the manuscript.

\section*{Funding}
The German Federal Ministry of Transport and Digital Infrastructure provides funding through the project “EasyRide” with grant number 16AVF2108H. \\
The authors remain responsible for all findings and opinions presented in the paper.

\bibliographystyle{plainnat}
\bibliography{bib}

\end{document}